\documentclass{article}%

\usepackage{amsmath,amsfonts,bm}

\def\1{\bm{1}}

\DeclareMathAlphabet{\mathsfit}{\encodingdefault}{\sfdefault}{m}{sl}
\SetMathAlphabet{\mathsfit}{bold}{\encodingdefault}{\sfdefault}{bx}{n}

\newcommand{\R}{\mathbb{R}}

\DeclareMathOperator*{\argmin}{arg\,min}

\usepackage{graphicx}%
\usepackage{multirow}%
\usepackage{amsmath,amssymb,amsfonts}%
\usepackage{amsthm}%
\usepackage{mathrsfs}%
\usepackage[title]{appendix}%
\usepackage{xcolor}%
\usepackage{textcomp}%
\usepackage{manyfoot}%
\usepackage{booktabs}%
\usepackage{algorithm}%
\usepackage{algorithmicx}%
\usepackage{algpseudocode}%
\usepackage{listings}%
\usepackage{makecell}
\usepackage{lmodern}
\usepackage{anyfontsize}
\usepackage{fullpage}
\usepackage{threeparttable}
\usepackage[colorlinks=true,linkcolor=black,anchorcolor=black,citecolor=black,filecolor=black,menucolor=black,runcolor=black,urlcolor=black]{hyperref}

\def\argmin{\mathop{\mathrm{arg\,min}}} %

\def\lim{\mathop{\mathrm{lim}}} %

\newcommand{\norm}[1]{\left\lVert#1\right\rVert}

\def\ebm{{\bm{e}}}

\def\xbm{{\bm{x}}}

\def\ybm{{\bm{y}}}

\def\wbm{{\bm{w}}}

\def\thetabm{{\bm{\theta}}}

\def\Abm{{\bm{A}}}
\def\Hbm{{\bm{H}}}
\def\Dbm{{\bm{D}}}
\def\Fbm{{\bm{F}}}
\def\Sbm{{\bm{S}}}
\def\Pbm{{\bm{P}}}

\def\xbmhat{{\widehat{\bm{x}}}}

\def\xbmbar{{\bar{\xbm}}}

\def\Tsf{{\mathsf{T}}}

\def\Dsf{{\mathsf{D}}}

\def\Isf{{\mathsf{I}}}

\def\R{\mathbb{R}}

\begin{document}

\title{Efficient Model-Based Deep Learning via Network Pruning and Fine-Tuning}
\date{}

\author{Chicago Y. Park\thanks{Denotes shared first authorship.},\ \ Weijie Gan\footnotemark[1],\ \ Zihao Zou, Yuyang Hu, Zhixin Sun, Ulugbek S. Kamilov \\
\small Washington University in St. Louis, MO, USA \\
\small \texttt{\{chicago, weijie.gan, zihao, h.yuyang, zhixin.sun, kamilov\}@wustl.edu}
}

\maketitle

\begin{abstract}
    Model-based deep learning (MBDL) is a powerful methodology for designing deep models to solve imaging inverse problems. MBDL networks can be seen as iterative algorithms that estimate the desired image using a physical measurement model and a learned image prior specified using a convolutional neural net (CNNs). The iterative nature of MBDL networks increases the test-time computational complexity, which limits their applicability in certain large-scale applications. Here we make two contributions to address this issue: First, we show how structured pruning can be adopted to reduce the number of parameters in MBDL networks. Second, we present three methods to fine-tune the pruned MBDL networks to mitigate potential performance loss. Each fine-tuning strategy has a unique benefit that depends on the presence of a pre-trained model and a high-quality ground truth. We show that our pruning and fine-tuning approach can accelerate image reconstruction using popular deep equilibrium learning (DEQ) and deep unfolding (DU) methods by 50\% and 32\%, respectively, with nearly no performance loss. This work thus offers a step forward for solving inverse problems by showing the potential of pruning to improve the scalability of MBDL. Code is available at \href{https://github.com/wustl-cig/MBDL\_Pruning}{\textcolor{magenta}{\text{https://github.com/wustl-cig/MBDL\_Pruning}}}.
\end{abstract}

\section{Introduction}\label{sec1}

The recovery of unknown images from noisy measurements is one of the most widely-studied problems in computational imaging and an instance of an \emph{inverse problem}. Conventional methods solve these problems by formulating optimization problems that consist of a data fidelity term enforcing consistency with the measurements and a regularizer imposing prior knowledge of the unknown images~\cite{Hu.etal2012,Elad.Aharon2006,Rudin.etal1992}. The focus in the area has recently shifted to methods based on \emph{deep learning (DL)}~\cite{Gilton.etal2020,Lucas.etal2018,McCann.etal2017}. A widely-used approach involves training a \emph{convolutional neural network (CNN)} to map the measurements directly to a high-quality reference in an end-to-end fashion~\cite{Kang.etal2017,Chen.etal2017,Wang.etal2016}.

Model-based deep learning (MBDL) has emerged as an alternative to traditional DL~\cite{Ongie.etal2020,Kamilov.etal2023,Monga.etal2021}. The key idea behind MBDL is to iteratively update images through operators that integrate the measurement models of the imaging systems and the learned CNNs. Notable examples of MBDL include \emph{plug-and-play (PnP)}~\cite{Venkatakrishnan.etal2013,Sreehari.etal2016}, \emph{regularization by denoising (RED)}~\cite{Romano.etal2017}, \emph{deep unfolding (DU)}~\cite{Schlemper.etal2018,Yang.etal2016,Hammernik.etal2018} and \emph{deep equilibrium models (DEQ)}~\cite{Gilton.etal2021,Heaton.etal2021}. Despite its superior performance, the iterative nature of MBDL also results in high computational cost during testing, limiting its applicability in large-scale applications. The computational complexity of MBDL arises from both the measurement models and the learned CNNs within the operators. Although several studies in MBDL have reduced the computational demand of the measurement models~\cite{Liu.etal2022,Wu.etal2020a,Sun.etal2019,Liu.etal2021a,Tang.Davies2020}, to the best of our knowledge, effort to mitigate the computational cost from the standpoint of the CNN priors remains unexplored.

\begin{figure*}[t]
\begin{center}
\includegraphics[width=1.\textwidth]{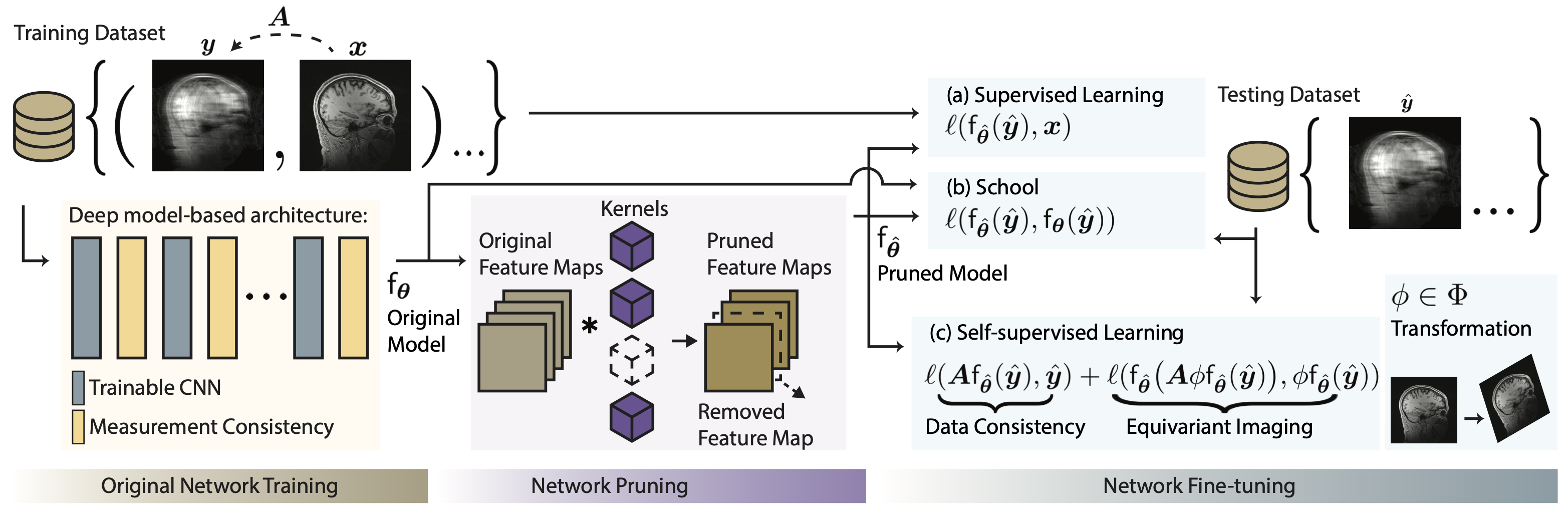}
\end{center}
\caption{An illustration of our proposed pipeline. Our proposed pipeline consists of two components (see Section~\ref{sec-method}): \emph{(a)} a structured pruning algorithm that physically removes CNN filters based on the group $\ell_1$-norm, and \emph{(b)} fine-tuning algorithms to minimize the performance loss between the pre-trained model and the pruned model. Each fine-tuning strategy has unique applicability depending on the presence of the pre-trained models and high-quality ground truth.}
\label{fig:method}
\vspace{-5pt}
\end{figure*}

In this paper, we bridge this gap by proposing a novel pipeline that combines network pruning and fine-tuning for obtaining more efficient MBDL networks for inverse problems. Our proposed pipeline uses the group $\ell_1$ -norm criteria to rank the importance of filters in the pre-trained CNN and then progressively eliminates filters from the least to the most important~\cite{Fang.etal2023}. Our proposed pipeline then integrates three distinct learning strategies for fine-tuning the pruned models depending on the availability of pre-trained models and ground truth: \emph{(a) supervised} penalizes the discrepancy between pruned model output and corresponding ground truth; \emph{(b) school} enforces consistency between pruned model output and that of the pre-trained model; \emph{(c) self-supervised} relies exclusively on testing dataset by using the losses of data fidelity and equivariant imaging~\cite{Chen.etal2021}. We show that our proposed pipeline can significantly increase the speed and efficiency of popular DU and DEQ models on two imaging problems: accelerated MRI and image super-resolution. Overall, our paper is the first to show that pruning and fine-tuning can significantly improve the scalability of MBDL with nearly no loss in imaging performance.

\section{Background}
\label{gen_inst}

\textbf{Inverse Problems.} Consider imaging inverse problems that aim to recover unknown images $\xbm\in\R^n$ from noisy measurements $\ybm\in\R^m$ characterized by a linear system
\begin{equation}
    \ybm = \Abm\xbm + \ebm\ ,
\end{equation}
where $\Abm\in\R^{m\times n}$ represents the measurement model of the imaging system, and $\ebm\in\R^m$ denotes an \emph{additive white Gaussian noise (AWGN)} vector. Due to the noise perturbation and ill-posedness (\emph{i.e.,} $m\ll n$), it is common to solve this problem by formulating an optimization problem 
\begin{equation}
    \xbmhat\in\argmin_{\xbm\in\R^n} f(\xbm) \text{ with } f(\xbm) = h(\xbm) + g(\xbm)\ ,
\end{equation}
where $h(\xbm)$ denotes the data-fidelity term that quantifies consistency with the measurements $\ybm$, and $g(\xbm)$ is the regularizer that imposes a prior on $\xbm$. For example, a widely-used data-fidelity term and regularizer in imaging are least-square $h(\xbm)=\frac{1}{2}\norm{\Abm\xbm-\ybm}_2^2$ and total variation $g(\xbm)=\tau\norm{\Dbm\xbm}_1$ where $\Dbm$ is the gradient operator, and $\tau$ is the trade-off parameter.

\textbf{DL and MBDL.} There is a growing interest in DL for solving inverse problems due to its excellent performance (see reviews in~\cite{Gilton.etal2020,Lucas.etal2018,McCann.etal2017}). A widely-used approach in this context trains a CNN to directly learn a regularized inversion that maps the measurements to high-quality reference~\cite{Kang.etal2017,Chen.etal2017,Wang.etal2016,Zhu.etal2018}. MBDL has emerged as a powerful DL framework for inverse problems by integrating the measurement models and learned CNNs (see also reviews in~\cite{Ongie.etal2020,Kamilov.etal2023,Monga.etal2021}). Notable examples of MBDL include \emph{plug-and-play (PnP)}~\cite{Venkatakrishnan.etal2013,Sreehari.etal2016}, \emph{regularization by denoising (RED)}~\cite{Romano.etal2017}, \emph{deep unfolding (DU)}~\cite{Schlemper.etal2018,Yang.etal2016,Hammernik.etal2018} and \emph{deep equilibrium models (DEQ)}~\cite{Gilton.etal2021,Heaton.etal2021}. PnP/RED refers to a family of algorithms that consider a CNN denoiser $\Dsf_\thetabm$ parameterized by $\thetabm$ as the imaging prior and then use $\Dsf_\thetabm$ in fixed-point iterations of some high-dimensional operators $\Tsf_\thetabm$. For example, the fixed-point iteration of the PnP proximal gradient method is formulated as 
\begin{equation}
\label{equ:pnp}
    \xbm^{k+1} = \Tsf_\thetabm(\xbm^{k}) = \Dsf_\thetabm(\xbm^k - \gamma\cdot\nabla h(\xbm^k))\ ,
\end{equation}
where $\gamma>0$ denotes the step size, and $k=0,..., K$. DU denotes a special end-to-end network architecture obtained by interpreting a \emph{finite} iteration of PnP/RED as different layers of the network. DEQ~\cite{Bai.etal2019,Bai.etal2020} is a recent approach that allows training \emph{infinite-depth}, \emph{weight-tied} networks by analytically backpropagating through the fixed points using implicit differentiation. Training DEQ for inverse problems is equivalent to training a weight-tied DU with infinite iterations.
To be specific, the forward pass of DEQ estimates a fixed-point $\xbmbar$ of the operator $\xbmbar=\Tsf_\thetabm(\xbmbar)$. The backward pass of DEQ updates $\Dsf_\thetabm$ by computing an implicit gradient of the training loss $\ell$ 
\begin{equation}
    \nabla \ell(\thetabm)= {\big(\nabla_{\thetabm}\Tsf(\xbmbar)\big)}^\Tsf {\big(\Isf-\nabla_{\xbm}\Tsf(\xbmbar)\big)}^{-\Tsf}\nabla \ell(\xbmbar) .
\end{equation}

MBDL has exhibited excellent performances in many imaging problems, such as MRI~\cite{Schlemper.etal2018,Gan.etal2020,Liu.etal2020,Sriram.etal2020,Cui.etal2023,Hammernik.etal2021,Hu.etal2022}, CT~\cite{Liu.etal2022,Liu.etal2021a,Adler.Oktem2018,Wu.etal2021a,He.etal2019}, and image restoration~\cite{Gilton.etal2021,Zhang.etal2021,Zhang.etal2020}. However, its iterative nature results in a high computational cost due to multiple CNN applications, limiting its use in large-scale or computation-constrained applications. While numerous studies have tackled the computational demand associated with the measurement models~\cite{Liu.etal2022,Wu.etal2020a,Sun.etal2019,Liu.etal2021a,Tang.Davies2020}, to the best of our knowledge, the reduction of the computational cost from the perspective of the CNN priors has not yet been explored.

\textbf{Network Pruning.} Network pruning denotes the process of eliminating weights, filters, or channels of a pre-trained network to obtain lightweight models (see also recent reviews in \cite{Ghimire2022,He2023,Cheng2023,Hoefler2021}). Pruning methods can be divided into unstructured pruning and structured pruning. Unstructured pruning \emph{virtually} masks out unimportant individual weights throughout the network. Since the unimportant weights are not removed physically, specialized software or hardware is required for computational acceleration in structural pruning \cite{Zhang2016,Parashar2017,Zhou2018_acc,Chen2019}. Structured pruning, on the other hand, \emph{physically} removes entire filters, channels, or layers, leading to faster inference without the need for any specialized hardware or software.
Many approaches have been proposed to identify the importance of the network filters prior to removing any of them, including those are based on \emph{(a)} certain criteria of the filters~\cite{He2018,Lin2020,Hu2016,Li2016}, such as $\ell_1$-norm~\cite{Li2016}, \emph{(b)} minimizing the reconstruction errors~\cite{Luo2017,Yu2018}, or \emph{(c)} finding the replaceable filters with similarity measurements~\cite{He2019,Zhou2018}.

\textbf{Fine-tuning Pruned Networks.} Following the pruning of the network, it is common to fine-tune the pruned model to minimize performance degradation (see also Section 2.4.6 in~\cite{Hoefler2021}). A widely-used strategy is to use the same amount of training data for the pre-trained models to retrain pruned models ~\cite{Lin2020,Hu2016,Li2016,Luo2017,Yu2018,Zhou2018,He2017,Lee2022,Han2015}. In imaging inverse problems, high-quality ground truth is commonly considered as the learning target for CNNs. However, ground truth data is not always available in practice, which limits the applicability of this fine-tuning approach in MBDL models. Other fine-tuning approaches include re-initialization of the pruned model's weight based on lottery-ticket-hypothesis~\cite{Frankle2018}, and \emph{knowledge distillation (KD)} that configures the pre-trained and pruned networks as a teacher-student pair~\cite{Lee2022,Hinton2015,Dong2019,Mirzadeh2020}. For example, KD in~\cite{Dong2019} proposes an auxiliary loss function to match the prediction of a pruned network and soft targets unpruned network.

\textbf{Self-supervised Deep Image Reconstruction.}
There is a growing interest in developing DL methods that reduce the dependence on the ground truth data (see recent reviews in~\cite{Akcakaya.etal2022a,tachella2023sensing,zeng2021review}).
Some widely-used strategies include \emph{Noise2Noise (N2N)}~\cite{Lehtinen.etal2018,Gan.etal2022},
\emph{Noise2Void (N2V)}~\cite{Krull.etal2019}, \emph{deep image prior (DIP)}~\cite{Ulyanov.etal2018}, \emph{compressive sensing using generative model (GSGM)}~\cite{bora2018ambientgan,gupta2021cryogan}, and \emph{equivariant imaging (EI)}~\cite{Chen.etal2021}.
In particular, EI assumes the set of ground truth is invariant to a certain group of transformations $\Phi$. The training loss of EI can then be formulated as
\begin{equation}
    \label{equ:ei}
    \ell_{\mathsf{EI}}(\thetabm) = \ell\big(\mathsf{f}_\thetabm(\Abm \xbmhat_\phi), \xbmhat_\phi\big) \text{ with } \xbmhat_\phi = \phi\ \mathsf{f}_\thetabm(\ybm)\ ,
\end{equation}
where $\mathsf{f}_\thetabm$ denotes the DL model, and $\phi\in\Phi$ is an instance of the transformation. The effectiveness of EI has been validated in a variety of imaging~\cite{Chen.etal2021,Chen.etal2022}, such as sparse-view CT and image inpainting. EI can also be integrated with another training loss, such as data-fidelity loss and adversarial loss~\cite{mao2017least}. 
As can be seen in the next section, we exploit EI to fine-tune our pruned model using exclusively the testing dataset.

\textbf{Our Contributions}: \emph{(1)} We propose the first application of network pruning and for MBDL models, aiming to reduce computational complexity at testing time. While the technique of network pruning has been extensively explored across a variety of tasks in computer vision, its potential has remained unexplored in the realm of imaging inverse problems; \emph{(2)} We develop three distinct fine-tuning methods to minimize the performance gap between pre-trained and pruned models. Each of these methods, to be detailed in the following section, is intuitive and holds practical applicability for inverse problems; \emph{(3)} We have conducted comprehensive experiments across various imaging problems, diverse MBDL methods, and different pruning ratios. Such extensive numerical validations represent a novel contribution, as they have not been performed in prior works.

\begin{figure*}[t]
\begin{center}
\includegraphics[width=1.\textwidth]{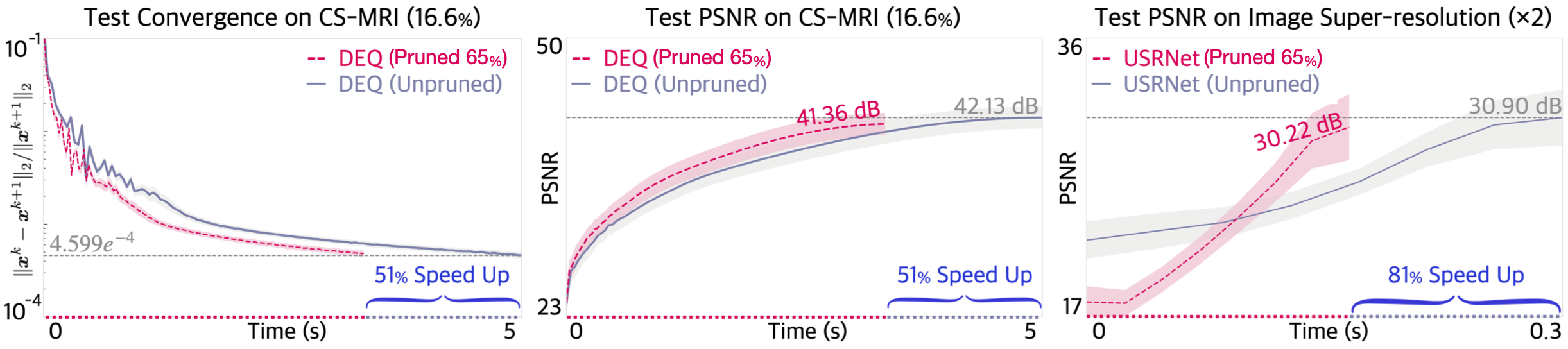}
\end{center}
\caption{Illustration of testing time evolution between MBDL models and the pruned variants using our proposed pipeline. \emph{Left:} evolution of the distance between two consecutive images; \emph{Middle and Right:} evolution of the testing PSNR values compared to the ground truth. We used the \emph{supervised} and \emph{school} strategies to fine-tune DEQ and USRNet, respectively. Note how models pruned by our proposed pipeline can significantly reduce the testing time while maintaining competitive performance.}
\label{fig:convergence_plot}
\end{figure*}

\section{Method}
\label{sec-method}

As illustrated in Figure~\ref{fig:method}, our proposed pipeline consists of a filter-pruning method and several fine-tuning methods. For the filter pruning method, we adopt DepGraph~\cite{Fang.etal2023} to identify layer dependencies and form \emph{layer groups} across the network. Let $\mathsf{f}_\thetabm$ denote the original unpruned model with $N$ layers. Let also $\mathsf{f}_{\thetabm,j}^-$ and $\mathsf{f}_{\thetabm,j}^+$ denote the input and the output of the $j$th layer $\mathsf{f}_{\thetabm,j}$, respectively. Consider two types of dependencies between $\mathsf{f}_{\thetabm,j}^-$ and $\mathsf{f}_{\thetabm,i}^+$ for all $i,j=1,\dots,N$: \emph{(a)} inter-layer dependency for $i\neq j$ indicates that $\mathsf{f}_{\thetabm,j}^-$ and $\mathsf{f}_{\thetabm,i}^+$ are topologically connected and correspond to the same intermediate features of the network, and 
\emph{(b)} intra-layer dependency for $i=j$ exists if and only if the mapping from $\mathsf{f}_{\thetabm,j}^-$ to $\mathsf{f}_{\thetabm,i}^+$ can be expressed as a diagonal matrix.
Another conceptual interpretation of intra-layer dependency is that $\mathsf{f}_{\thetabm,j}^+$ and $\mathsf{f}_{\thetabm,i}^-$ \emph{share the same pruning scheme}~\cite{Fang.etal2023}.
For example, consider a convolutional layer ($\mathsf{Conv}$). Consider a filter ${\bm K}\in\R^{K_\textsf{in}\times K_\textsf{out}\times H \times W}$ in a $\mathsf{Conv}$, where $K_\textsf{in}$ denotes the number of input channels, $K_\textsf{out}$ is the number of output channels, and $H\times W$ represents the kernel size. Pruning the input of a $\mathsf{Conv}$ necessitates pruning the kernel along the $K_\textsf{in}$ dimension. Conversely, pruning the output demands alterations along the $K_\textsf{out}$ dimension. The difference in pruning schemes for the input and the output of $\mathsf{Conv}$ indicates the absence of an intra-layer dependency. 
DepGraph examines all inputs and output pairs to compute their dependencies (see also Algorithm 1 in~\cite{Fang.etal2023}).

The layer groups across the network can then be constructed based on the identified dependencies (also refer to Algorithm 2 in \cite{Fang.etal2023}).
Each group must adhere to the following conditions:
\emph{(a)} the group can be represented as a connected graph, where the nodes are the layers within the group, and the edges denote the topological connections between the input and output of two layers (\emph{i.e.,} inter-layer dependency);
\emph{(b)} the hidden layers (\emph{i.e.,} non-edge layers) within the group must exhibit intra-layer dependencies.
Note that the inter-layer and intra-layer dependencies ensure that the prunable dimensions, such as the number of feature map channels in CNN, are identical across different layers in the same group. For example, consider a sample network of $\{\mathsf{Conv}_1\rightarrow\mathsf{BN}_1\rightarrow\mathsf{Conv}_2\rightarrow\mathsf{BN}_2\}$, where $\rightarrow$ denotes topological connection, and $\mathsf{BN}$ is the batch normalization layer.
The resulting layer groups include $\{\mathsf{Conv}_1, \mathsf{BN}_1, \mathsf{Conv}_2\}$ and $\{\mathsf{Conv}_2, \mathsf{BN}_2\}$. The prunable dimension of the first and second groups match the output channel of $\mathsf{Conv}_1$ (or the input channel of $\mathsf{Conv}_2$) and the output channel of $\mathsf{Conv}_2$, respectively.

We use the group $\ell_1$-norm to evaluate the importance of parameters in each layer group. To be specific, consider a layer group with $M$ layers with $i=1,...,M$ denoting the $i$th layer within the group. Let also $\wbm_k^i$ be the filters of the $k$th prunable dimension in the $i$th layer for $k=1,...,K$. The group $\ell_1$ norm vector of a layer group is formulated as $\bm{\alpha} = [\alpha_1, \cdots, \alpha_k]$ where
\begin{equation}
    {\alpha_k=\frac{1}{M}\sum^M_{i=1}\norm{\wbm_k^i}_1}, \quad k=1,\dots,K\ .
\end{equation}
For each layer group, we compute the group $\ell_1$ norm and subsequently select a subset of the smallest $\alpha_k$ according to a pre-defined pruning ratio. Parameters associated with this subset are then pruned.

Let $\mathsf{f}_{\hat{\thetabm}}$ be the pruned model. We consider three distinct fine-tuning algorithms to minimize the performance loss between $\mathsf{f}_{\hat{\thetabm}}$ and $\mathsf{f}_{{\thetabm}}$. Each fine-tuning method has unique applicability depending on the presence of the unpruned model
and high-quality ground truth.

\begin{table*}[t]
\caption{Quantitative evaluation of MBDL models pruned by our proposed pipeline with different fine-tuning strategies in CS-MRI at the sampling rate of $16.6\ \%$. The `Speed Up' column indicates how much faster each pruned model is compared to the unpruned model in terms of prediction time for processing 100 input sequences.}
\label{tb:cs-mri-x6}
\begin{center}
\setlength{\textwidth}{1.75pt}
\setlength\tabcolsep{5pt}
{\tiny%
\begin{tabular}{p{0.6cm} p{0.7cm} p{0.7cm} p{0.7cm} p{0.7cm} p{0.7cm} p{0.1cm} p{0.7cm} p{0.7cm} p{0.7cm} p{0.7cm} p{0.55cm} ccp{.5cm}}
\toprule
\multirow{2}{*}{Network} & \multicolumn{1}{c}{\multirow{2}{*}{\makecell{Pruning\\Ratio}}} & \multicolumn{4}{c}{PSNR (dB)} & & \multicolumn{4}{c}{SSIM (\%)} & \multirow{2}{*}{\makecell{Speed \\ Up}} & \multirow{2}{*}{\makecell{\#Params \\ (millions)}} \\ \cmidrule{3-6} \cmidrule{8-11}
                      & \multicolumn{1}{c}{}                               & \emph{Supervised}     & \emph{School}     & \emph{Self-supervised}  & \emph{No-tuning} &  & \emph{Supervised}     & \emph{School}     & \emph{Self-supervised}   & \emph{No-tuning}  &                       &                             \\
\midrule
\multirow{5}{*}{DEQ}  & $0\ \%$   &  \multicolumn{4}{c}{42.13 {}}   &      &     \multicolumn{4}{c}{{98.7}}                      & $\times$1.00                    & 1                   \\ \cmidrule{2-13}
                      & $10\ \%$   &  42.02       &  41.85        &   40.37   &  39.47     & & 98.6      &  98.6       &  98.2   &  97.9   & $\times$1.03 &        0.88      \\
                      & $20\ \%$   & 42.02      & 41.69       &  38.79   &  37.69  & & 98.7      & 98.6      & 97.7    &  97.3  &       $\times$1.06             & 0.79                             \\
                      & $35\ \%$   & 41.94        & 41.32        & 36.14     &  35.41   & &     98.6  & 98.5       & 96.5     &  96.1    & $\times$1.17                  & 0.64                             \\
                      & $65\ \%$   & 41.36       & 39.99        & 34.54   &  33.26      & &   98.5      & 98.2        & 95.7      &  94.9     & $\times$1.51                   & 0.35   \\  
\midrule
\multirow{5}{*}{VarNet}  & $0\ \%$   &  \multicolumn{4}{c}{39.25}   &      &     \multicolumn{4}{c}{97.7}                    & $\times $1.00             & 19.63                    \\ \cmidrule{2-13}
                      & $10\ \%$   &  39.16       &  38.86        &  39.11   &  29.34   & & 97.7      &  97.6       & 97.4   &  89.8           & $\times$1.00  &  17.57             \\
                      & $20\ \%$   & 39.06      & 38.71       &  38.59   &  26.39  & & 97.7    & 97.5       & 97.3   &  81.4               & $\times$1.06 & 15.77                            \\
                      & $35\ \%$   & 38.85       & 38.54       & 37.46   &  22.81     & &    97.6   & 97.4       & 96.9    &  73.4     & $\times$1.12                  &    12.48                         \\
                      & $65\ \%$   & 38.12       & 37.36        & 34.17     &  20.65    & &    97.3     & 96.7        & 95.3     &  64.5   & $\times$1.32                      & 6.99  \\     
\midrule
\multirow{5}{*}{E2EVar}  & $0\ \%$   &  \multicolumn{4}{c}{44.24 }   &      &     \multicolumn{4}{c}{99.2 }                       &   $\times$1.00   &  20.12              \\ \cmidrule{2-13}
                      & $10\ \%$   &  44.18       & 43.76        &   40.12  &  26.66   & &  99.2      & 99.1       & 97.7    &  80.8          &   $\times$1.01                &   18.05       \\
                      & $20\ \%$  & 44.17       & 43.24        &  39.91  &  25.89   & & 99.2      & 99.0       & 97.6     &  83.7    &         $\times$1.05              &   16.26   \\
                      & $35\ \%$   & 43.71       & 42.66       & 39.28     &  24.38   & &     99.1   & 98.9        & 97.3   &  75.6         &   $\times$1.09            &   12.96           \\
                      & $65\ \%$   & 42.62        & 41.58         & 37.82    &  23.46    & &    98.8      & 98.5         & 96.9     &  70.2                 &  $\times$1.24       &  7.47  \\     
\bottomrule
\end{tabular}
}
\end{center}
\vspace{-10pt}
\end{table*}

\begin{figure*}[t]
\begin{center}
\includegraphics[width=1.\textwidth]{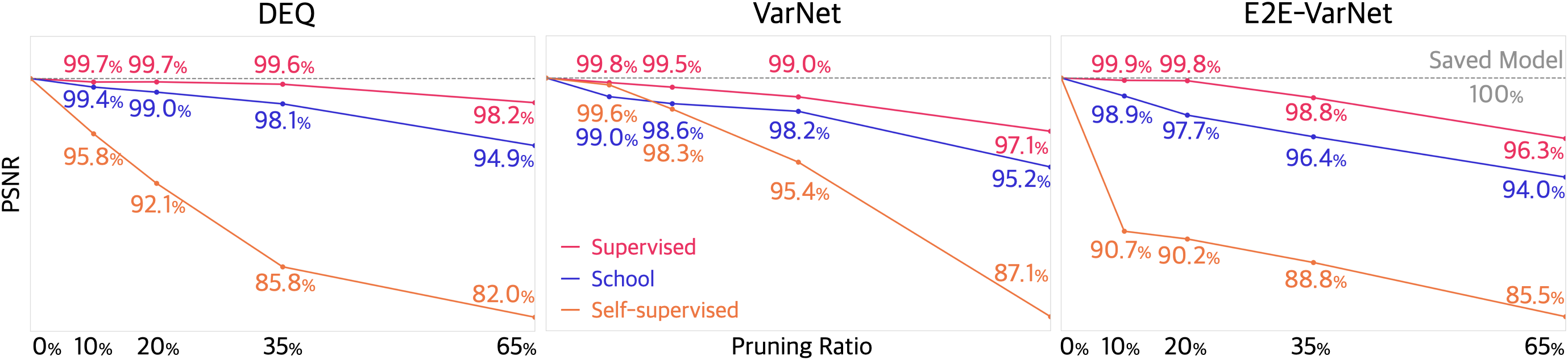}
\end{center}
\caption{Degradation PSNR percentage of pruned MBDL models compared to the unpruned model in different pruning ratios and different fine-tuning strategies. These results correspond to experiments of CS-MRI at the sampling rate of $16.6\ \%$. Note how \emph{supervised} fine-tuning method can reduce $65\%$ parameters while maintaining less than $4\%$ PSNR degradation.}
\label{fig:cs-mri-degrad-x6}
\end{figure*}

\emph{(a)} \textbf{supervised (SV):} SV considers a training set consisting of pairs of measurements and ground truth $\{\ybm_i, \xbm_i\}_{i=1}^N$, where $N$ denotes the total number of training samples. The loss function of SV minimizes the difference between the reconstruction of the pruned model and ground truth
\begin{equation}
    \ell_\mathsf{sv}(\hat{\thetabm}) = \sum_{i = 1}^N \ell(\mathsf{f}_{\hat{\thetabm}}(\ybm_i),\ \xbm_i)\ .
\end{equation}
This training strategy is also the most common scheme for training CNNs for imaging inverse problems starting from scratch. Despite its effectiveness (see Sec.~\ref{sec:exp-mri}), SV relies on a collection of high-quality ground truth, which might not always be available in practice. 

\emph{(b)} \textbf{school (SC):} SC considers a testing set of unseen measurements $\{\ybm_j\}_{j = 1}^M$ where $M$ denotes the total number of testing samples. The loss function of SC penalizes the discrepancy between the outputs of the pruned model and those of the unpruned model
\begin{equation}
    \ell_\mathsf{sc}(\hat{\thetabm}) = \sum_{j=1}^M \ell(\mathsf{f}_{\hat{\thetabm}}(\ybm_j),\ \mathsf{f}_{{\thetabm}}(\ybm_j))\ .
\end{equation}
SC can be seen as a new branch of KD in imaging, where we transfer the ``knowledge'' from the unpruned model (\emph{teacher}) to the pruned model (\emph{student}). The reason why it is called school follows. Unlike SV, SC only requires access to the pre-trained model.

\emph{(c)} \textbf{self-supervised (SS):} Compared to SC, SS considers a particular case where the pre-trained models are unavailable for some reason (e.g., privacy). The key idea behind SS is to fine-tune $\mathsf{f}_{\hat{\thetabm}}$ using exclusively the testing set $\{\ybm_j\}_{j=1}^M$. Let $\Phi$ denote a certain group of transformations. The loss function of SS can be formulated as
\begin{equation}
\label{equ:ss}
    \ell_\mathsf{ss}(\hat{\thetabm}) = \sum_j^M \underbrace{\ell(\Abm_j\mathsf{f}_{\hat{\thetabm}}(\ybm_j),\  \ybm_j)}_\textsf{Data Fidelity} + \underbrace{\ell\big(\mathsf{f}_{\hat{\thetabm}}(\Abm_j \xbmhat_{\phi,j}), \xbmhat_{\phi,j}\big)}_\textsf{Equivariant Imaging}\ .
\end{equation}
where $\xbmhat_{\phi,j} = \phi_j\ \mathsf{f}_{\hat{\thetabm}}(\ybm_j)$, and $\phi_j\sim\Phi$ is a transformation sampled randomly from $\Phi$.

\begin{table*}[t]
\tiny
\caption{Quantitative evaluation of MBDL models pruned by our proposed pipeline with different fine-tuning strategies in CS-MRI at the sampling rate of $12.5\ \%$. The `Speed Up' column follows the same definition as in Table \ref{tb:cs-mri-x6}}
\label{tb:cs-mri-x8}
\begin{center}
\setlength{\textwidth}{1.75pt}
\setlength\tabcolsep{5pt}
\begin{tabular}{p{0.6cm} p{0.7cm} p{0.7cm} p{0.7cm} p{0.7cm} p{0.7cm} p{0.1cm} p{0.7cm} p{0.7cm} p{0.7cm} p{0.7cm} p{0.55cm} ccp{.5cm}}
\toprule
\multirow{2}{*}{Network} & \multicolumn{1}{c}{\multirow{2}{*}{\makecell{Pruning\\Ratio}}} & \multicolumn{4}{c}{PSNR (dB)} & & \multicolumn{4}{c}{SSIM (\%)} & \multirow{2}{*}{\makecell{Speed \\ Up}} & \multirow{2}{*}{\makecell{\#Params \\ (millions)}} \\ \cmidrule{3-6} \cmidrule{8-11}
                      & \multicolumn{1}{c}{}                               & \emph{Supervised}     & \emph{School}     & \emph{Self-supervised}  & \emph{No-tuning}   &  & \emph{Supervised}     & \emph{School}     & \emph{Self-supervised}   & \emph{No-tuning}  &                       &                             \\
\midrule
\multirow{5}{*}{DEQ}  & $0\ \%$   &  \multicolumn{4}{c}{38.07 }   &      &     \multicolumn{4}{c}{96.9}                     &      $\times$1.00                    & 1                   \\  \cmidrule{2-13}
                      & $10\ \%$   &  38.06      &  37.87       &  36.13  &  35.54      & &  96.9     &  96.9      & 96.2   &  95.8       & $\times$1.03 &         0.88     \\
                      & $20\ \%$   & 38.07      & 37.77       &  34.94  &  34.26      & & 96.9     & 96.8      & 95.6     &  95.1        &    $\times$1.06             & 0.79                             \\
                      & $35\ \%$   & 37.68        & 37.42        & 32.54    &  32.13       & &   96.8   & 96.7        & 93.9    &  93.3           & $\times$1.17                  & 0.64                             \\
                      & $65\ \%$   & 37.13        & 36.33        & 30.26    &  28.99       & &    96.5      & 96.2         & 90.4    &  87.8          & $\times$1.51                   & 0.35   \\   
\midrule
\multirow{5}{*}{VarNet}  & $0\ \%$   &  \multicolumn{4}{c}{36.17}   &      &     \multicolumn{4}{c}{96.3}    & $\times $1.00             & 19.63                    \\ \cmidrule{2-13}
                      & $10\ \%$   &  36.04       &  36.04       &  35.80  &  29.36       & &  96.3      &  96.2       & 95.9   &  87.1        & $\times$1.00  &  17.57                             \\
                      & $20\ \%$   & 36.06       & 35.89        & 35.36  &  26.49        & & 96.3      & 96.2       & 95.8    &  81.6                     & $\times$1.06 & 15.77                            \\
                      & $35\ \%$   & 35.80        & 35.55        & 34.69   &  23.05         & &     96.1   & 95.9       & 95.2 &  72.5      & $\times$1.12                  &    12.48                         \\
                      & $65\ \%$   & 35.30        & 34.07         & 31.13   &  21.55         & &    95.7      & 94.9         & 92.6  &  67.5     & $\times$1.32                      & 6.99  \\   
\midrule
\multirow{5}{*}{E2EVar}  & $0\ \%$   &  \multicolumn{4}{c}{40.41}   &      &     \multicolumn{4}{c}{98.0}                                                &   $\times$1.00   &  20.12               \\ \cmidrule{2-13}
                      & $10\ \%$   & 40.31       & 40.00        &  37.31  &  24.37     & &  98.0      &  97.9       & 96.4    &  81.4               &   $\times$1.01                &   18.05       \\
                      & $20\ \%$   & 40.23      & 39.72        &  37.30   &  23.06     & & 98.0      & 97.9      & 96.2  &  77.5        &             $\times$1.05              &   16.26   \\
                      & $35\ \%$   & 40.14        & 39.30       & 35.74   &  22.60       & &    97.9  & 97.6       & 95.6     &  69.6       &          $\times$1.09            &   12.96           \\
                      & $65\ \%$   & 38.93        & 38.43         & 34.62     &  22.27       & &  97.4      & 97.1        & 95.0   &  67.6        &    $\times$1.24       &  7.47  \\    
\bottomrule
\end{tabular}
\end{center}
\vspace{-10pt}
\end{table*}

\begin{figure*}[t]
\begin{center}
\includegraphics[width=1.\textwidth]{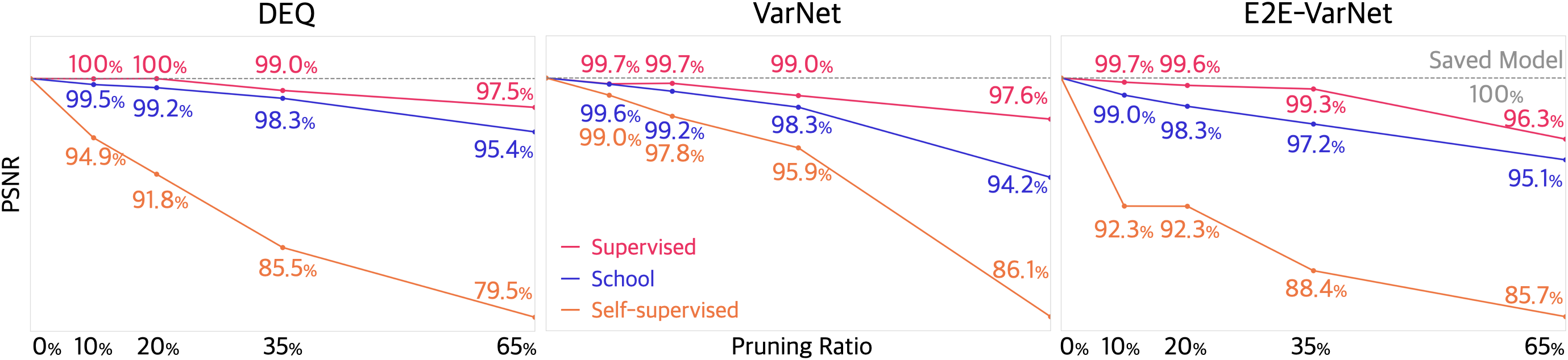}
\end{center}
\caption{Degradation PSNR percentage of pruned MBDL models compared to the unpruned model in different pruning ratios and different fine-tuning strategies. These results correspond to experiments of CS-MRI at the sampling rate of $12.5\ \%$. Note how \emph{supervised} fine-tuning method can reduce $65\%$ parameters while maintaining less than $4\%$ PSNR degradation.}
\label{fig:cs-mri-degrad-x8}
\end{figure*}

\section{Numerical Results}
\label{sec-exp}

The goal of our experiments was to validate the effectiveness of our proposed pipeline on different imaging problems, different MBDL networks, and different pruning ratios.
We used three fine-tuning methods introduced in ~\ref{sec-method}. For the \textit{supervised} fine-tuning method, we used the same training data as the pre-training of the unpruned model, which consists of $90\%$ of the entire dataset. In contrast, for the \textit{school} and \textit{self-supervised} fine-tuning methods, we used only $5\%$ of the undersampled dataset (i.e., without ground truth), which was not included in the supervised training data.
We tested our proposed pipeline on two imaging problems: CS-MRI and image super-resolution.
We used \emph{ peak signal-to-noise ratio (PSNR)} and \emph{structural similarity index measure (SSIM)} for quantitative evaluation.
We pre-defined pruning ratios of $5\%$, $10\%$, $20\%$, and $40\%$, resulting in actual pruning ratios of $10\%$, $20\%$, $35\%$, and $65\%$ since the layers dependent on the pruned layers are also removed. (see Sec.~\ref{sec-method}).
All pre-trained models were trained in a supervised learning manner to ensure their optimal performance.
We implemented $\Phi$ in \eqref{equ:ss} as a set of rotations.
We pre-trained and fine-tuned the MBDL models by using the Adam optimizer with the learning rate being $10^{-5}$. We conducted all experiments on a machine equipped with an AMD EPYC 7443P 24-Core Processor and 4 NVIDIA RTX A6000 GPUs.

\subsection{Compressed Sensing MRI}
\label{sec:exp-mri}
The measurement model of CS-MRI consists of a set of complex measurement operators depending on a set of receiver coils $\{\Sbm_i\}$. For each coil, we have $\Abm_i=\Pbm\Fbm\Sbm_i$, where $\Fbm$ is the Fourier transform, $\Pbm$ denotes the diagonal sampling matrix, and $\Sbm_i$ is the diagonal matrix of sensitivity maps. We used T2-weighted MR brain acquisitions of 165 subjects obtained from the validation set of the fastMRI dataset~\cite{Knoll2020} as the fully sampled measurement for simulating measurements. We obtained reference coil sensitivity maps from the fully sampled measurements using ESPIRiT~\cite{Uecker.etal2014}. These 165 subjects were split into 145, 10, and 10 for training, validation, and testing, respectively. We followed ~\cite{Knoll2020} to retrospectively undersample the fully sampled data using 1D Cartesian equispaced sampling masks with 10\% auto-calibration signal (ACS) lines. We conducted our experiments for the sampling rate of $16.7\%$ and $12.5\%$.

We tested our proposed pipeline on DEQ and two DU models: VarNet~\cite{Hammernik.etal2018}, and E2E-VarNet~\cite{Sriram.etal2020}. We implemented DEQ with forward iteration as in \eqref{equ:pnp} and EDSR~\cite{Lim.etal2017} as the CNN architecture. We ran the forward-pass of DEQ with a maximum number of iterations of 100 and the stopping criterion of the relative norm difference between iterations being less than $10^{-4}$. The implementations of VarNet and E2E-VarNet were from their official repository\footnote{\url{https://github.com/facebookresearch/fastMRI}}. The difference between E2E-VarNet and VarNet is that E2E-VarNet has an additional coil sensitivity estimator compared to VarNet. Noted that our proposed pipeline did not prune this estimator for E2E-VarNet. We also used the estimated coil sensitivity map for \emph{self-supervised} fine-tuning.
The training of unpruned models for E2E-VarNet and VarNet took around seven days, and that for DEQ was ten days.

Table~\ref{tb:cs-mri-x6} shows quantitative evaluations of MBDL models pruned by our proposed pipeline in MRI at a sampling rate of $16.6\%$. Figure~\ref{fig:cs-mri-degrad-x6} illustrates the PSNR degradation percentage of the pruned model compared to the unpruned network at the same sampling rate. Figure~\ref{fig:cs-mri-degrad-x6} demonstrates that the \emph{supervised} fine-tuning strategy is highly effective, resulting only up to $4\%$ PSNR degradation to achieve approximately $65\%$ fewer parameters and up to $\times$1.51 speed up. Both Figure~\ref{fig:cs-mri-degrad-x6} and Table~\ref{tb:cs-mri-x6} also indicate that one can achieve a $\times$1.06 speed up in testing time with virtually no cost in PSNR degradation by removing $20\%$ of parameters. Moreover, Figure~\ref{fig:cs-mri-degrad-x6} highlights that the \emph{school} fine-tuning method can achieve competitive performance against \emph{supervised} fine-tuning across different pruning ratios. Table~\ref{tb:cs-mri-x6} further shows that DEQ can achieve higher speed up than DU models under the same pruning ratio, which we attribute to its large number of forward iterations. Table~\ref{tb:cs-mri-x8} and Figure~\ref{fig:cs-mri-degrad-x8} present similar evaluations as in Table~\ref{tb:cs-mri-x6} and Figure~\ref{fig:cs-mri-degrad-x6}, but at a sampling rate of $12.5\%$, with consistent observations.

\begin{table*}[t]
\tiny
\caption{Quantitative evaluation of MBDL models pruned by our proposed pipeline with different fine-tuning strategies in CS-MRI at the sampling rate of $12.5\ \%$ with noise corresponding
to input SNR = 9 dB.
Note how school and self-supervised fine-tuning methods can perform comparably to supervised fine-tuning despite requiring only $1/18$th of the data used by the supervised method. The `Speed Up' column follows the same definition as in Table \ref{tb:cs-mri-x6}}
\label{tb:cs-mri-x8_noiseintense}
\begin{center}
\setlength{\textwidth}{1.75pt}
\setlength\tabcolsep{5pt}
\begin{tabular}{p{0.6cm} p{0.7cm} p{0.7cm} p{0.7cm} p{0.7cm} p{0.7cm} p{0.1cm} p{0.7cm} p{0.7cm} p{0.7cm} p{0.7cm} p{0.55cm} ccp{.5cm}}
\toprule
\multirow{2}{*}{Network} & \multicolumn{1}{c}{\multirow{2}{*}{\makecell{Pruning\\Ratio}}} & \multicolumn{4}{c}{PSNR (dB)} & & \multicolumn{4}{c}{SSIM (\%)} & \multirow{2}{*}{\makecell{Speed \\ Up}} & \multirow{2}{*}{\makecell{\#Params \\ (millions)}} \\ \cmidrule{3-6} \cmidrule{8-11}
                      & \multicolumn{1}{c}{}                               & \emph{Supervised}     & \emph{School}     & \emph{Self-supervised}  & \emph{No-tuning}   &  & \emph{Supervised}     & \emph{School}     & \emph{Self-supervised}   & \emph{No-tuning}  &                       &                             \\
\midrule
\multirow{3}{*}{DEQ}  & $0\ \%$   &  \multicolumn{4}{c}{29.73 }   &      &     \multicolumn{4}{c}{88.7}                     &      $\times$1.00                    & 1                   \\  \cmidrule{2-13}
                      & $10\ \%$   &  29.72      &  29.60       &  29.36  &  27.93      & &  89.0     &  89.1      &  88.5   &  82.2       & $\times$1.03 &         0.88     \\
                      & $65\ \%$   & 29.68        & 29.45        & 27.71    &  22.28       & &   88.5   & 88.5        & 86.2    &  61.7           & $\times$1.17                  & 0.64                             \\
\midrule
\multirow{3}{*}{VarNet}  & $0\ \%$   &  \multicolumn{4}{c}{28.88}   &      &     \multicolumn{4}{c}{89.4}    & $\times $1.00             & 19.63                    \\ \cmidrule{2-13}
                      & $10\ \%$   &  28.86       &  28.82       &  27.68  &  26.41       & &  89.4      &  89.2       & 87.9   &  82.8        & $\times$1.00  &  17.57                             \\
                      & $65\ \%$   & 28.55        & 27.99        & 27.22   &  18.64         & &     89.4   & 87.4       & 86.9 &  55.1      & $\times$1.12                  &    12.48                         \\
\midrule
\multirow{3}{*}{E2EVar}  & $0\ \%$   &  \multicolumn{4}{c}{29.96}   &      &     \multicolumn{4}{c}{91.0}                                                &   $\times$1.00   &  20.12               \\ \cmidrule{2-13}
                      & $10\ \%$   & 29.84       & 29.88        &  28.25  &  21.00     & &  90.9      &  90.1       & 89.0    &  72.5               &   $\times$1.01                &   18.05       \\
                      & $65\ \%$   & 29.66        & 29.57       & 24.92   &  18.83       & &    90.6  & 90.3       & 79.3     &  56.4       &          $\times$1.09            &   12.96           \\
\bottomrule
\end{tabular}
\end{center}
\vspace{-5pt}
\end{table*}

Notably, Table~\ref{tb:cs-mri-x8_noiseintense} evaluates a more challenging setting by adding extra noise to the existing measurement noise in real data, with the added noise corresponding to an input SNR of 9 dB.
Despite this increased difficulty, school and self-supervised fine-tuning methods, which use only undersampled test data, remain effective and comparable to supervised fine-tuning methods that rely on ground truth from the entire training dataset.

\subsection{Image Super-Resolution}
We consider the measurement model of form $\Abm=\Sbm\Hbm$, where $\Hbm\in\R^{n\times n}$ is the blurring matrix, and $\Sbm\in\R^{m\times n}$ denotes the standard $d$-fold down-sampling operator with $d^2=n/m$. We evaluated our proposed pipeline on the CSDB68 dataset. We conducted our experiments for down-sampling factors of $2$ and $3$. We followed~\cite{Zhang.etal2020} to experiment with eight different Gaussian blur kernels and four motion kernels. We tested our proposed pipeline on a DU model, USRNet~\cite{Zhang.etal2020}, in the \emph{school} fine-tuning method to simulate the circumstance where only a pre-trained network is accessible. We used the pre-trained model provided by the official repository\footnote{\url{https://github.com/cszn/USRNet}}.

\begin{figure*}[t]
\begin{center}
\vspace{10pt}
\includegraphics[width=\textwidth]{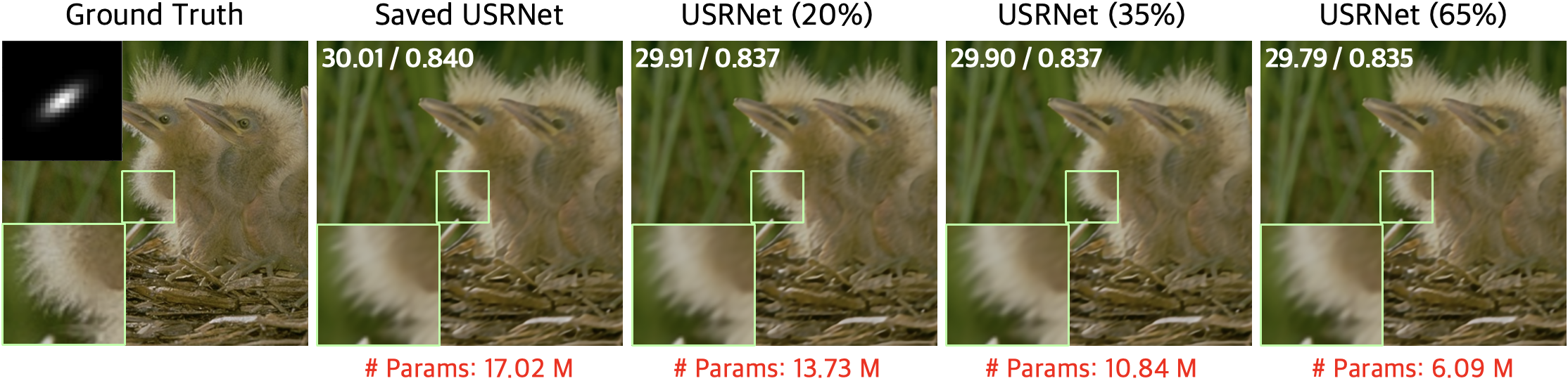}
\end{center}
\caption{Visual results of USRNet and its pruned variants with the \emph{school} fine-tuning strategy at scale of $\times$3. Note how USRNet has been pruned more than $65\%$ parameters but kept similar visual quality compared to the unpruned model.}
\label{fig:usrnet}
\end{figure*}

Table~\ref{tb:usrnet} shows the quantitative evaluation of USRNet pruned by our proposed pipeline with \emph{school} fine-tuning strategy in image super-resolution at the scale of $\times$2 and $\times$3. This table highlights that the \emph{school} fine-tuning strategy can achieve $\times$1.81 speed up while maintaining less than $1\%$ degradation in both PSNR and SSIM values. Figure~\ref{fig:usrnet} shows visual results of USRNet and its pruned variants in image super-resolution at the scale of $\times$3. Figure~\ref{fig:usrnet} demonstrates that the pruned models can achieve qualitatively competitive performance compared to the unpruned network.

\begin{table*}[t]
    \vspace{0.1in}
    \caption{Quantitative evaluation of USRNet pruned by our proposed pipeline with \emph{school} fine-tuning strategy in image super-resolution at the scale of $\times$2 and $\times$3. The `Speed Up' column follows the same definition as in Table \ref{tb:cs-mri-x6}}
    \vspace{0.1in}
    \label{tb:usrnet}
    \centering
        \tiny
        \renewcommand\arraystretch{1}

        \begin{threeparttable}
        \begin{tabular}{cccccccccc}
        \toprule
        \multirow{2}{*}{Scale} & \multicolumn{1}{c}{\multirow{2}{*}{Pruning Ratio}} & \multicolumn{2}{c}{PSNR (dB)} & & \multicolumn{2}{c}{SSIM (\%)} & \multirow{2}{*}{Time (ms)} & \multirow{2}{*}{Speed Up} & \multirow{2}{*}{\makecell{\#Params \\ (millions)}} \\ \cmidrule{3-4} \cmidrule{6-7}
                              & \multicolumn{1}{c}{}                               & \emph{School} & \emph{Degrad.} $\%$   &  & \emph{School}     &   \emph{Degrad.} $\%$   &                    &                             \\
        \midrule
        \multirow{4}{*}{$\times 2$}  & $0\ \%$   &  29.96    & $100.0\ \%$ & &  86.5   & $100.0\ \%$ & 272.2  &  $\times$1.00     & 17.02               \\ \cmidrule{2-10}
                              & $10\ \%$  & 29.81  & $99.5\ \%$ & & 86.2   & $99.6\ \%$ & 262.8 & $\times$1.04 & 15.31 \\
                              & $20\ \%$  & 29.81  & $99.5\ \%$ & & 86.1   & $99.5\ \%$ & 248.9 & $\times$1.09 & 13.73 \\
                              & $35\ \%$  & 29.80  & $99.5\ \%$ & & 86.1  & $99.5\ \%$ & 241.8 & $\times$1.13 & 10.84 \\
                              & $65\ \%$  & 29.70  & $99.1\ \%$ & & 86.0  & $99.4\ \%$ & 150.3 & $\times$1.81 & 6.09 \\
        \midrule
        \multirow{4}{*}{$\times 3$}  & $0\ \%$   &  27.56    & $100.0\ \%$ & &  79.0   & $100.0\ \%$ &     272.2     & $\times$1.00  & 17.02                \\ \cmidrule{2-10}
                              & $10\ \%$  & 27.43 & $99.5\ \%$ & & 78.5  & $99.3\ \%$ & 262.8 & $\times$1.04 & 15.31 \\
                              & $20\ \%$  & 27.43 & $99.5\ \%$ & & 78.5  & $99.3\ \%$ & 248.9 & $\times$1.09 & 13.73 \\
                              & $35\ \%$  & 27.42 & $99.5\ \%$ & & 78.5  & $99.3\ \%$ & 241.8 & $\times$1.13 &  10.84 \\
                              & $65\ \%$  & 27.32 & $99.1\ \%$ & & 78.2  & $98.9\ \%$ & 150.3 & $\times$1.81 & 6.09 \\
        \bottomrule
        \end{tabular}
            \begin{tablenotes}
                \item The ``\emph{Degrad. \%}'' columns denote degradation percentage of PSNR values of pruned models compared to that of pruned models. Note how \emph{school} fine-tuning method can gain 1.81$\times$ speed up with less than $1\%$ performance degradation.
            \end{tablenotes}
        \end{threeparttable}
        \vspace{-10pt}
\end{table*}

\section{Conclusion}

This work presents the first pipeline to reduce the test-time computational complexity of model-based deep learning \emph{through neural network pruning and fine-tuning}. Our proposed pipeline employs group $\ell_1$-norm to identify the significance of CNN weights, pruning them in ascending order of importance. We propose three distinct fine-tuning strategies to minimize the performance deviation between pruned and pre-trained models. Each of these fine-tuning methods possesses unique applications, contingent on the availability of high-quality ground truth and a pre-trained model: \emph{(a) supervised} strategy minimizes the discrepancy between the output of the pruned model and the corresponding ground truth; \emph{(b) school} ensures consistency between the outputs of the pruned and the pre-trained models; \emph{(c) self-supervised} exclusively relies on the testing dataset, leveraging data fidelity and equivariant imaging losses.

We evaluated the efficacy of our proposed pipeline through applications in compressed sensing MRI and image super-resolution, employing several MBDL models. The experimental results in MRI demonstrate that: the \emph{(a) supervised} strategy can realize up to $\times$1.51 speed up in testing time by eliminating $65\%$ of parameters, with less than $4\%$ degradation in testing PSNR values; it can also attain $\times$1.06 speed up with negligible performance cost; \emph{(b) school} can achieve competitive performance against \emph{supervised}, with less than $3\%$ PSNR degradation across different pruning ratios and MBDL models; \emph{(c) self-supervised} can outperform equivalently parameterized models trained from scratch using the same loss function. The results in image super-resolution further corroborate the effectiveness of our proposed pipeline on the \emph{school} fine-tuning method. Future directions for this research include testing our proposed pipeline with alternative approaches to ranking the importance of CNN weights, and exploring different losses to enhance the self-supervised fine-tuning methods.

\section*{Acknowledgments}
This paper is partially based upon work supported by the NSF CAREER award under grants CCF-2043134.

\end{document}